\documentclass[aps,twocolumn,pre,superscriptaddress,showpacs,showkeys,nofootinbib ]{revtex4-1}

\usepackage{amsmath}
\usepackage{amssymb}
\usepackage{graphicx}                   % for insert file eps
\usepackage{hyperref}                   % for hyper refrences
\usepackage{amsmath,amssymb}            % for insert extra math
\usepackage{epstopdf}
\usepackage{subcaption}					% for sub-figures
\usepackage{color}
\usepackage{nicefrac, xfrac}
\definecolor{orange}{rgb}{1,0.5,0}
\definecolor{brown}{rgb}{0.65, 0.16, 0.16}
\definecolor{phlox}{rgb}{0.87, 0.0, 1.0}
\usepackage[dvipsnames,svgnames,x11names]{xcolor}
\usepackage[markup=underlined]{changes}
\definechangesauthor[color=BrickRed]{ML}

\usepackage{enumitem}
\usepackage{todonotes}
\graphicspath{{figs/}}					% the path of image folder 

\begin{document}

\title{Simulating Cumulus Clouds based on Self-Organized Criticality}

\author{J. Cheraghalizadeh}
\affiliation{Department of Physics, University of Mohaghegh Ardabili, P.O. Box 179, Ardabil, Iran}

\author{M. Lukovi\'c}
\affiliation{Computational Physics, IfB, ETH Zurich, Stefano-Franscini-Platz 3, CH-8093 Zurich, Switzerland}
\affiliation{Cellulose and Wood Materials, Empa Swiss Federal Laboratories for Materials Science and Technology, CH-8600 Dübendorf, Switzerland}

\author{M. N. Najafi}
\affiliation{Department of Physics, University of Mohaghegh Ardabili, P.O. Box 179, Ardabil, Iran}
\email{morteza.nattagh@gmail.com}

\begin{abstract}
Recently it was shown that self-organized criticality is an important ingredient of the dynamics of cumulus clouds (Physical Review E, 103(5), p.052106, 2021). Here we introduce a new algorithm to simulate cumulus clouds in two-dimensional square lattices, based on two important facts: the cohesive energy of wet air parcels and a sandpile-type diffusion of cloud segments. The latter is realized by considering the evaporation/condensation of air parcels in various regions of the cloud, which enables them to diffuse to the neighboring regions. The results stemming from this model are in excellent agreement with the observational results reported in the above-cited paper, where the exponents have been obtained for the two-dimensional earth-to-sky RGB images of clouds. The exponents that are obtained at the lowest condensation level in our model are consistent with the observational exponents. We observed that the cloud fields that we obtain from our model are fractal, with the outer perimeter having a fractal dimension of $D_f=1.25\pm 0.01$. Furthermore, the distributions of the radius of gyration and the loop length follow a power-law function with exponents $\tau_r=2.3\pm 0.1$ and $\tau_l=2.1\pm 0.1$, respectively. The loop Green function is found to be logarithmic with the radius of gyration of the loops following the observational results. The winding angle statistic of the external perimeter of the cloud field is also analyzed, showing an exponent in agreement with the fractal dimension, which may serve as the conformal invariance of the system. 
\end{abstract}


\maketitle

\section{Introduction}

Convective clouds such as cumulus clouds (member of the cumuliform clouds) develop in unstable air due to the buoyancy resulting from water vapor, supercooled water droplets, or ice crystals, depending upon the ambient conditions. The upward movement of warm air results in the formation of cumulus clouds at a threshold called the lifting condensation level (LCL), where the relative humidity reaches 100\%, at which the nucleation process starts around various nuclei present in the air. The cumulus clouds are less than 2 $km$ in altitude unless they develop more vertically, in which case they become cumulus congestus and may appear in the form of lines or clusters. While they produce little or no precipitation, they are often precursors to the formation of other types of clouds, e.g. cumulonimbus, when influenced by weather factors such as instability, moisture, and temperature gradient. This complex dynamics, resulting in cotton-like low-level cumulus clouds with flat bases, includes many degrees of freedom and dynamical parameters, such as convection, condensation and evaporation, diffusion, and self-organized criticality, that have recently been discussed \cite{PhysRevE.103.052106} (see \cite{guichard2017short} for a good reference). A complete treatment for this complex non-equilibrium system includes numerous dynamical parameters and physics including the Navier-Stocks equation, self-organized propagation of the wet air parcels, thermodynamic of mixing of air parcels, condensation and evaporation physics, placing this system into the list of complex systems. More suitable treatment needs a reasonable downfolding of the degrees of freedom without harming the key physical processes, which is the aim of the present paper. \\

Fractal or multi-fractal geometry \cite{mandelbrot1982fractal} is a powerful tool for classifying clouds in terms of their formation conditions. Some properties of clouds, like self-affinity and scaling features, have already been determined from satellite images\cite{lovejoy1982area,austin1985small}. Several observables, like the relation between cloud area and perimeter \cite{lovejoy1982area,austin1985small, austin1985small, chatterjee1994fractal,von2011first,malinowski1993surface,batista2016classification}, rainfall time series analysis \cite{olsson1993fractal}, the distribution function of geometrical quantities \cite{benner1998characteristics,rodts2003size,yano1987self,gotoh1998fractal} and the nearest neighbor spacing \cite{joseph1990nearest}, were shown to represent scaling behaviors. Following these investigations together with multi-fractal analysis \cite{lovejoy1991multifractal,lovejoy1987functional,cahalan1989fractal,gabriel1988multifractal}, the clouds can be placed into universality classes based on cloud morphology \cite{sengupta1990cumulus,lovejoy1990multifractals,tessier1993universal,pelletier1997kardar} and their field statistics.

In addition to satellite and ground-to-sky photos, cloud properties can be studied and analyzed using simulation methods. One can assume three major categories for cloud simulations. One is based on turbulence and includes atmospheric fluid dynamics models \cite{miyazaki2001method,wang2010shape,yano1996fractality}, models based on fractal geometry (or the scaling) in turbulent flow \cite{malinowski1993surface,richardson1926atmospheric,hentschel1984relative}, stochastic models based on cascade processes \cite{tessier1993universal,schertzer1987physical}, and large eddy simulations for cumulus convective transport \cite{siebesma2000anomalous,zhao2005life}. The second category is based on computer graphics techniques for cloud simulation \cite{gardner1985visual,cianciolo1993cumulus,bouthors2004modeling} and cellular automata \cite{nagel1992self}. The third category comprises models based on heuristic methods such as the  Kardar-Parisi-Zhang (KPZ) equation \cite{pelletier1997kardar}, the midpoint displacement algorithm \cite{cai2013cumulus} and the diamond square algorithm describing the fractal properties of cloud edges \cite{lohmann2017simulating}.

\begin{figure*}
	\centering
	\begin{subfigure}{0.4\textwidth}\includegraphics[width=\textwidth]{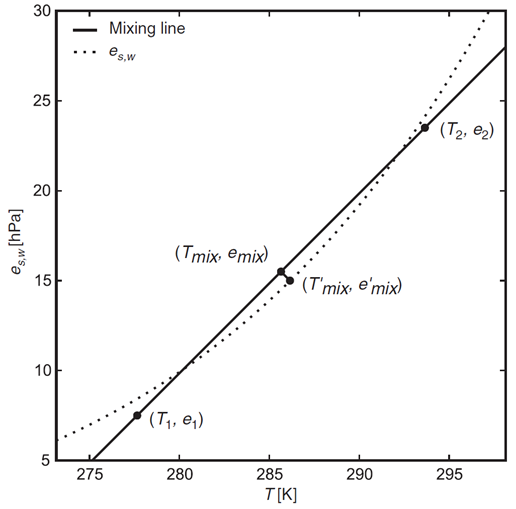}
		\caption{}
		\label{fig:mixture}
	\end{subfigure}
	\centering
	\begin{subfigure}{0.4\textwidth}\includegraphics[width=\textwidth]{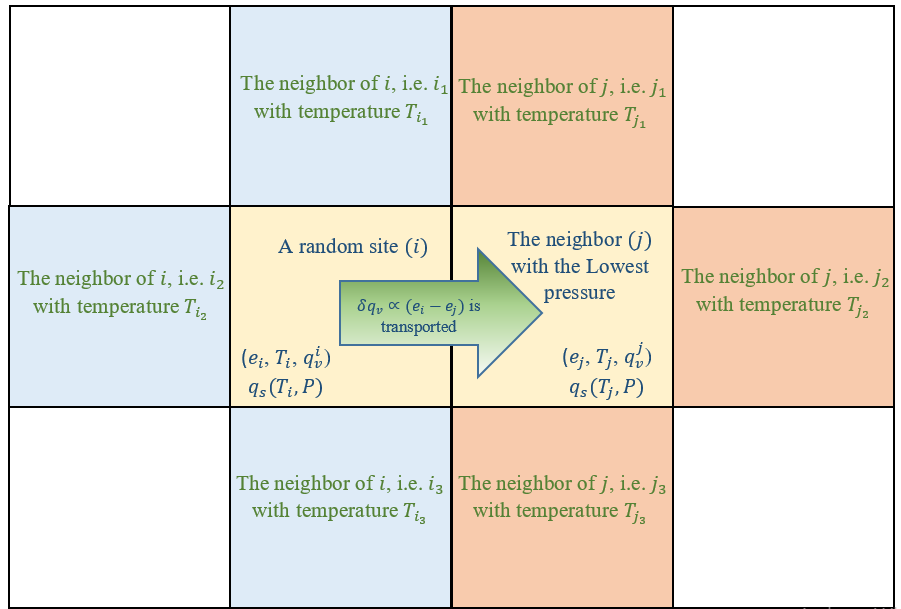}
		\caption{}
		\label{fig:SchematicAlgorithm}
	\end{subfigure}
\caption{(Color online) (a) Mixing of two air parcels. When the partial pressure rises beyond the saturation threshold, it condenses and the new temperature can be obtained by following the straight line from the \textit{mix} value (bold line) to the saturation curve (dotted curve). (b) Schematic representation of the algorithm outlined in the text. }
	\label{Schematic}
\end{figure*}

Furthermore, the studies mentioned above also show that clouds can be described by scaling relations and their related exponents, which are robust under most atmospheric conditions. These circumstances suggest that clouds organize themselves into critical states without the need for a tuning parameter.

The presence of Self-organized criticality (SOC) in the atmosphere and in clouds was first detected by analyzing precipitation, where the time series of the size of rain was analyzed for which a SOC mechanism was proposed \cite{peters2006critical}. A recent study uncovers the SOC state of clouds directly by analyzing earth-to-sky images of cumulus clouds and using a combination of the Navier-Stokes equation, diffusion equations, and a coupled map lattice (CML) as well \cite{PhysRevE.103.052106}. In that study, Schramm-Loewner evolution (SLE) analysis was used to show that cloud boundaries belong to the sandpile universality class, i.e. $c = -2$ conformal field theory (CFT).

The present paper aims to introduce an optimal two-dimensional lattice model with two key factors: The cohesive energy between the droplets leading to a contentious Ising model, and a sandpile model for the diffusion of air parcels within a horizontal plane. We will use the facts presented above to introduce and develop an algorithm based on the Metropolis Monte Carlo method to capture the physics of air parcel movement in clouds. To this end, we use the cohesive energy of water droplets and the physics governing the evaporation/condensation of air parcels in different cloud regions. Each part of the algorithm has been designed carefully not to miss any of the internal degrees of movement of the air parcels of the clouds. The exponents that are obtained at LCL are equivalent to the observational exponents.

We start our discussion by outlining the details of our algorithm in the following section, where we also explain how the simulations were done. Subsequently, we analyze our results and compare them with those of Ref.\cite{PhysRevE.103.052106} in section \ref{result} as well. We close the paper with a short conclusion.

\section{The Thermodynamics of Cumulus clouds; our model}

 When cumulus clouds grow into the congestus or cumulonimbus clouds, they are more probable to precipitate. The height of the cloud (from its bottom to its top) depends on the temperature profile of the atmosphere. These clouds form via atmospheric convection as warm air heated by the surface begins to rise, resulting in a decrease in temperature and rise in humidity. At the LCL, condensation to the wet-adiabatic phase starts. The released latent heat (due to condensation) warms up further the air parcel, resulting in further convection. At the LCL, the nucleation process starts around various nuclei present in the air. Details regarding the process of raindrop formation and rainfall can be found for example in~\cite{langmuir1948production} by Langmuir. Although the liquid water density within a cumulus cloud changes with height above the cloud base \cite{stommel1947entrainment} (for the non-precipitating clouds the concentration of droplets ranges from 23 to 1300 droplets per cubic centimeter \cite{warner1969microstructure}), the density can be thought of as being approximately constant throughout the cloud. The height of the cumulus clouds depends on the amount of moisture in the atmosphere, and  humid air will generally result in a lower cloud base. In stable air conditions in which their vertical growth is not that high, they are considered to be effectively two-dimensional.\\

The basic atmospheric properties (or fields, to be more precise) that play a vital role in cloud formation are the specific humidity $q_v$, the temperature $T'$, and the atmosphere pressure $P$. The relative humidity $q_v$ is defined as the mass of water vapor $m_v$ per unit mass of moist air $m_m$. The \textit{saturation of specific humidity} ($q_s(T',P)$) is defined as the specific humidity above which the system changes phase to liquid, which is approximated to be~\cite{lohmann2016}
\begin{equation}
q_s(T',P)= \frac{\epsilon A}{P}\exp\left[-B/T'\right]
\label{Eq:saturation}
\end{equation}
where $A=2.53\times 10^{11}\text{Pa}$, $B=5420\text{K}$, and $\epsilon\equiv\frac{M_w}{M_d}=0.622$, $M_d=28.96 \text{g}\ \text{mol}^{-1}$ is the molecular weight of dry air, and $M_w=18 \text{g}\ \text{mol}^{-1}$ is the molecular weight of water. In such a two-phase case, we need to work with \textit{partial pressures}. The partial water vapour pressure is denoted by $e$. We can make use of the fact that the water vapour pressure is much smaller than the atmospheric pressure ($e\ll P$) so that the partial water pressure can be approximated as~\cite{lohmann2016}
\begin{equation}
e=\frac{P}{\epsilon}q_v.
\end{equation}
As an air parcel moves upwards (rises), the cloud is formed for the first time at the LCL, which is identified by the relation 
\begin{equation}
q_v=q_s(T',P).
\end{equation}
In order to work with numbers that are suitable for making the simulations easier and faster, we cast Eq.~\ref{Eq:saturation} into an adimensional form
\begin{equation}
	Q_s=\frac{1.574}{P(\text{atm})}\exp\left[-\frac{B'-9T\ln 10}{T} \right], 
\end{equation}
where $T=\frac{T'}{100}$, $B'=54.2$, $Q_s=10^3q_s$, and $Q_v=10^3q_v$. As an application, let us approximate the physical parameters at LCL for cumulus clouds. For the adiabatic evolution of air parcels, one expects that 
\begin{equation}
T'(h)\approx T'(0)-9.8(\frac{K}{km})\times h,
\end{equation}
where $h$ is the altitude. Therefore, knowing that the typical altitude of LCL for the cumulus clouds is $2000\text{m}$, one expects that at the LCL: $P^{\text{LCL}}=0.93 atm$, $T^{\text{LCL}}\approx 2.7$ and $Q_v^{\text{LCL}}=Q_s^{\text{LCL}}\approx 3.17$.

\subsection{Air Parcels and the Cohesive Energy}

In this paper, we propose an internal mechanism responsible for the observations presented in Ref. \cite{PhysRevE.103.052106}, i.e. the self-organized critical behavior with some critical exponents for fractal cumulus clouds. An \textit{air parcel} is an imaginary volume of air to which the basic dynamic and thermodynamic properties of air are assigned. It starts from the sea level and rises. It condenses and becomes part of the cloud when it reaches the LCL. The cloud starts forming only at the LCL and not below this point. Furthermore, Furthermore, on the one hand, an air parcel must be large enough to contain many molecules; on the other hand, it also must be small enough so that its macroscopic properties be approximately uniform inside it. The motion of air parcels with respect to the surrounding atmosphere is supposed not to induce significant compensatory movements in the environment. One could visualize such an air parcel as having dimensions of some tens of centimeters in each direction. An air parcel is assumed to be thermally insulated from its environment so that its temperature changes adiabatically as it rises or sinks~\cite{lohmann2016}.

\begin{figure*}
	\centering
	\begin{subfigure}{0.16\textwidth}\includegraphics[width=\textwidth]{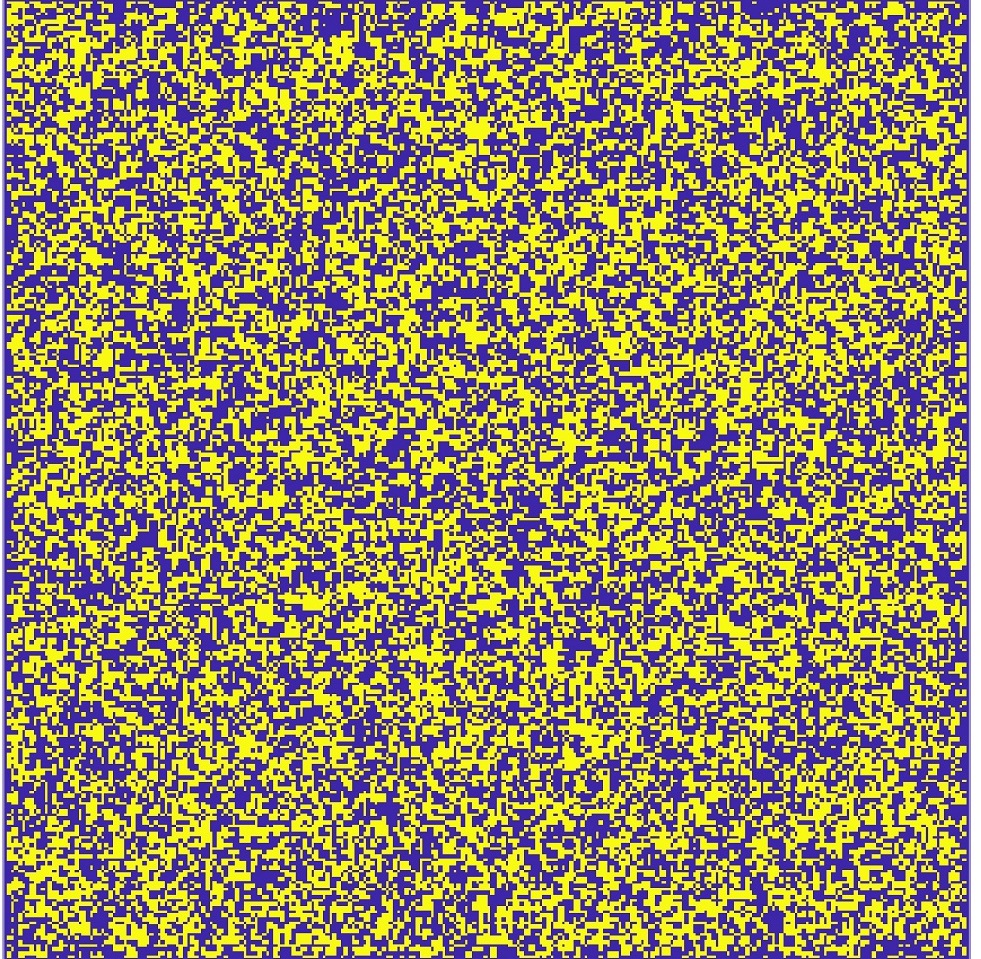}
		\caption{}
		\label{fig:t1}
	\end{subfigure}
	\centering
	\begin{subfigure}{0.16\textwidth}\includegraphics[width=\textwidth]{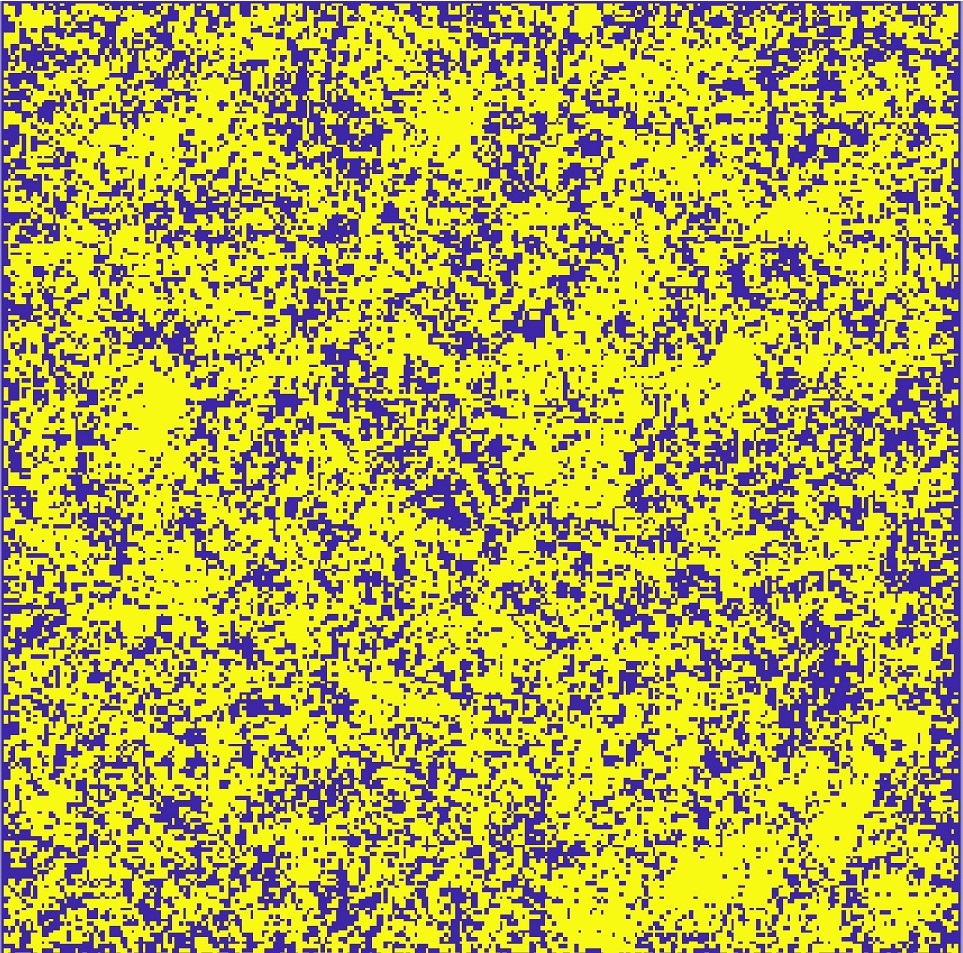}
		\caption{}
		\label{fig:t2}
	\end{subfigure}
	\begin{subfigure}{0.16\textwidth}\includegraphics[width=\textwidth]{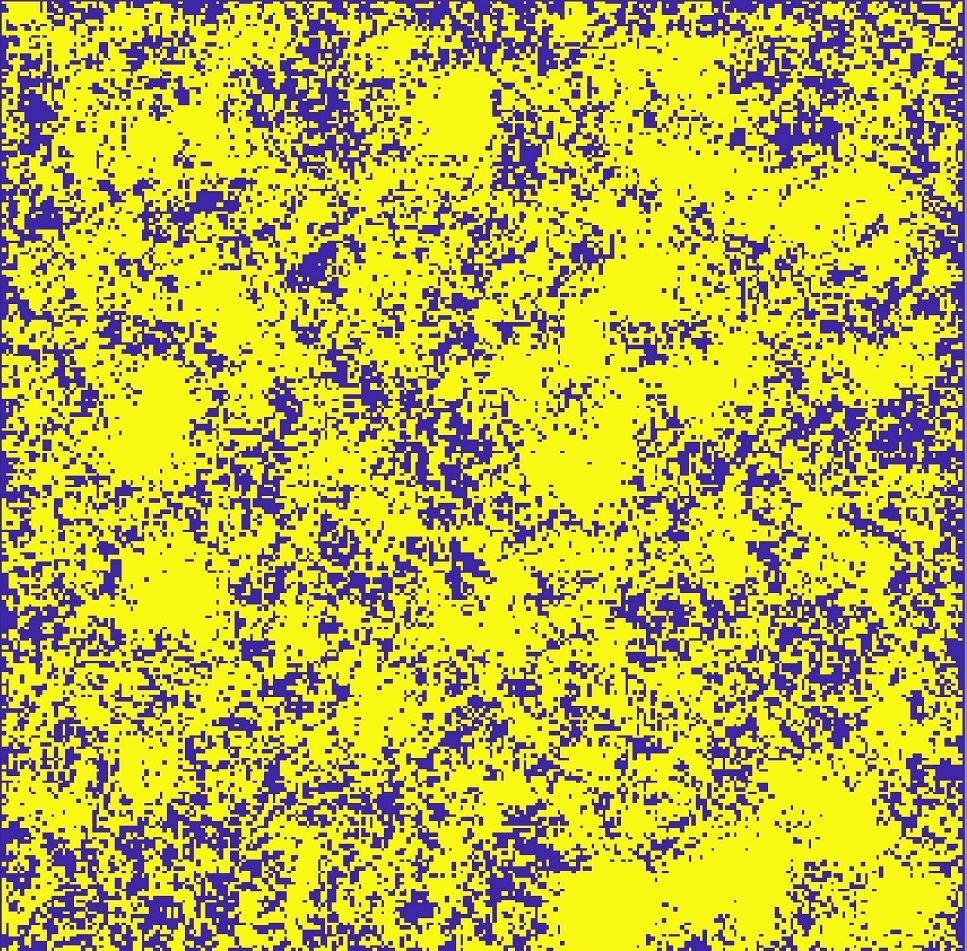}
		\caption{}
		\label{fig:t3}
	\end{subfigure}
	\begin{subfigure}{0.16\textwidth}\includegraphics[width=\textwidth]{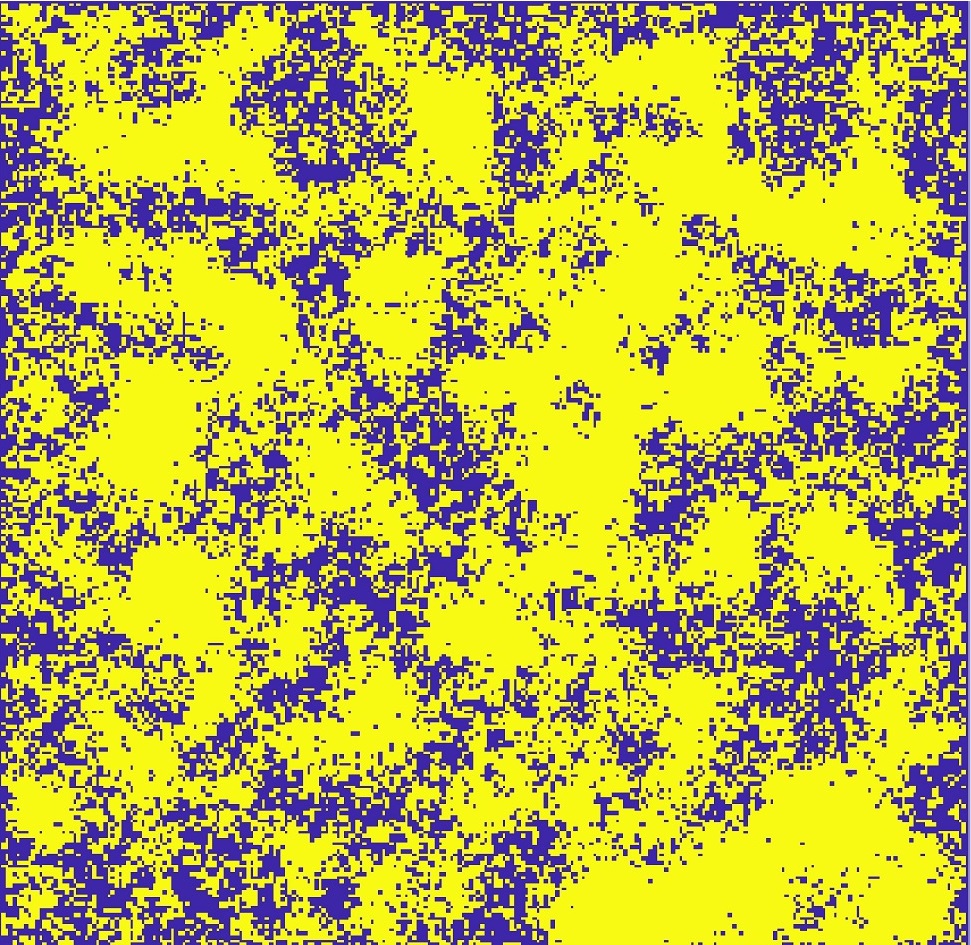}
		\caption{}
		\label{fig:t4}
	\end{subfigure}
	\begin{subfigure}{0.16\textwidth}\includegraphics[width=\textwidth]{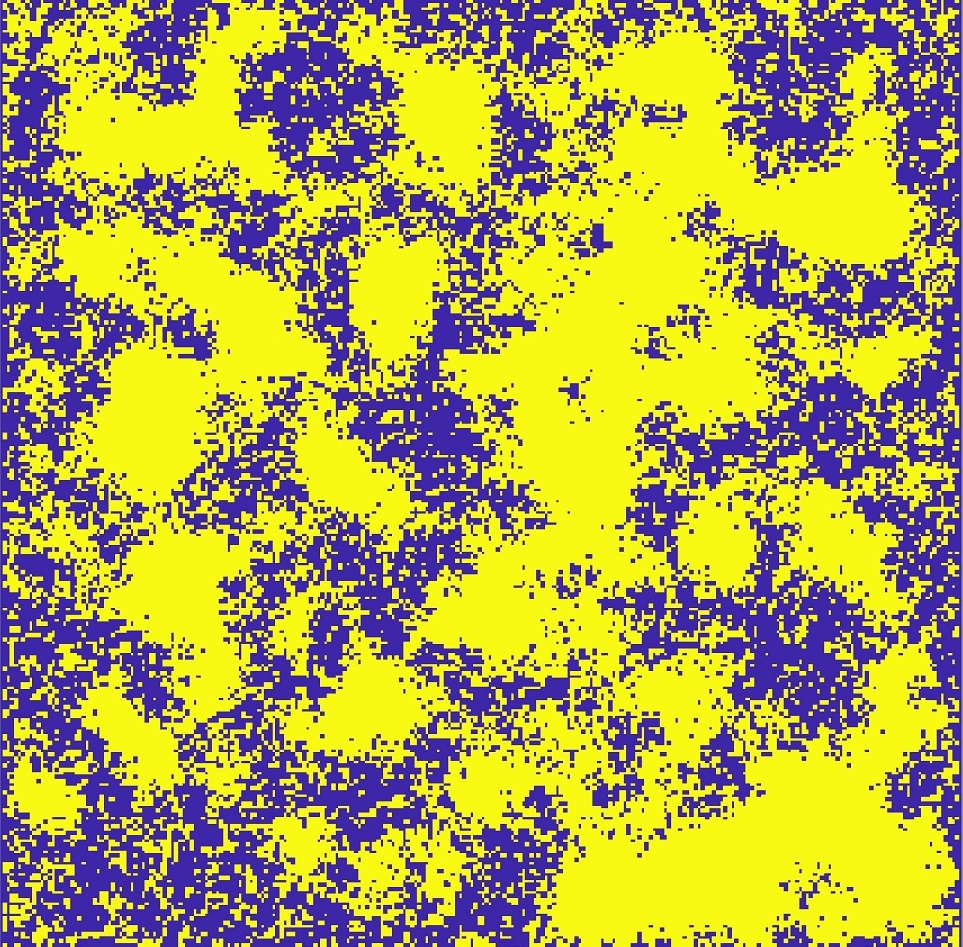}
		\caption{}
		\label{fig:t5}
	\end{subfigure}
	\begin{subfigure}{0.16\textwidth}\includegraphics[width=\textwidth]{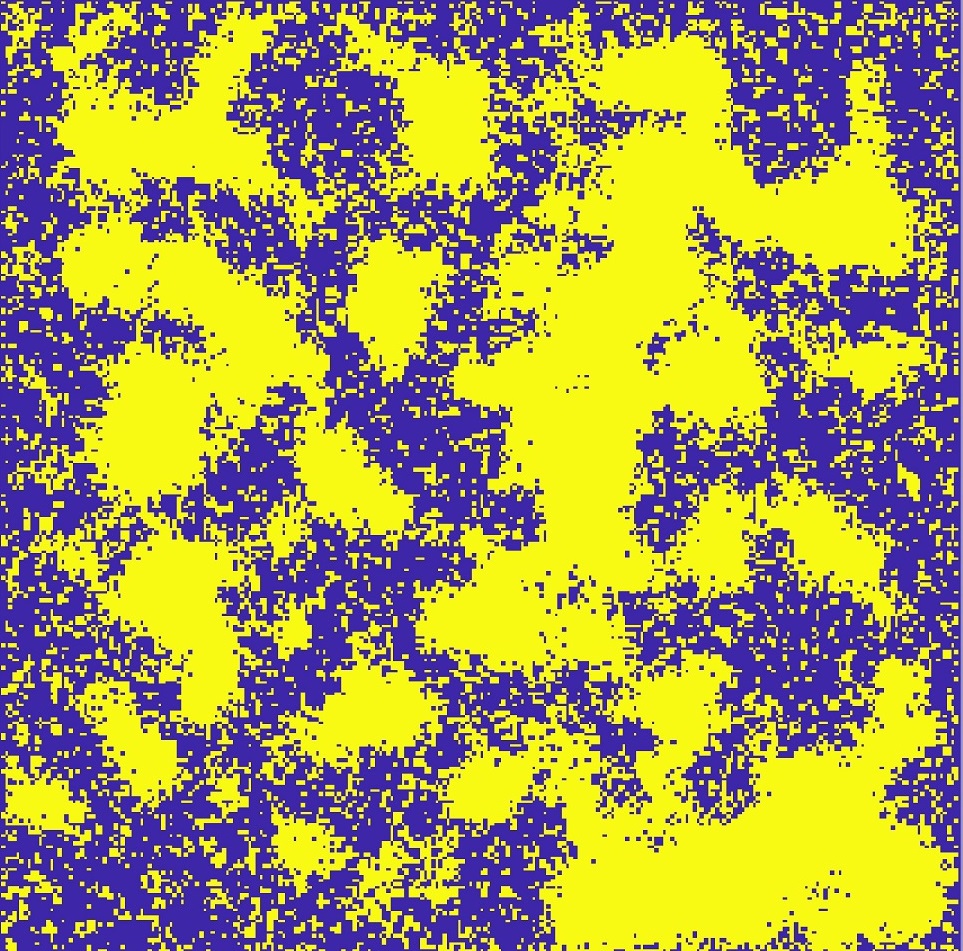}
		\caption{}
		\label{fig:t6}
	\end{subfigure}
	\caption{(Color online) The samples resulting from the simulation as a function of Monte Carlo steps for $T = 2.7$, $Q_s = 3.17$, and $P = 0.93$ atm. In (f) the system has reached the steady state. Each pixel is one parcel. Yellow=cloud.}
	\label{fig:sh1}
\end{figure*}

In addition to the atmospheric properties, the cohesive energy of water droplets per molecule, $J_M$, plays an important role in the pattern formation of clouds. More precisely, without the tendency of water droplets to cohere to each other (given by the cohesive energy), the clouds look like uncorrelated random media in which the segments diffuse. In our model, we add the possibility od cohering of the water droplets during the formation of clouds. The cohesive energy is nearly $41$ kJ/mol which is $0.464 eV$ per molecule, giving rise to $\frac{J_M}{kT}\approx \frac{46.4\text{eV}}{kT'}$. To quantify the cohesive energy and incorporate it to our lattice model, we consider two $d$-dimensional boxes ($d=2$ in our paper) of air parcels in contact, containing $N_1$ and $N_2$ water droplets. Then the number of droplets on the interface, for say the first box, is proportional to $N_1^{\frac{d-1}{d}}$, so that the cohesive energy can be approximated by:
\begin{equation}\label{eq:cohesive_energy}
\text{Cohesive Energy}\sim -J\left[ \left(N_1N_2\right)^{\frac{d-1}{d}}+\frac{1}{2}\left( N_1^2+N_2^2\right)\right],
\end{equation}
where $J\propto J_M$ is the cohesive energy between two water droplets. Equation \ref{eq:cohesive_energy} is our direct contribution and the starting point for a lattice-based dynamic model for cloud formation and dissipation. The minus sign guarantees the absorbing nature of the interactions. The first term describes the cohesive energy at the boundary between the two parcels (boundary effect). Assuming that the density is uniform, the number of the pairwise interactions at the boundary scales as the first term and the number of droplets is proportional to $N^{\frac{d-1}{d}}$. The second term in the equation describes the bulk/total (bulk term) energy inside the two connecting parcels and it is proportional to the number of pairwise interactions in the bulk. The resulting cloud phase is determined by the competition between the cohesive energy (equation \ref{eq:cohesive_energy}) between parcels and the entropic force (described in the algorithm below), which requires that the parcels become well-mixed and the cloud sparse. In other words, two phenomena are at play in the entire system, the adhesive effect between two parcels and the diffusion of each parcel within the total cloud system. When two parcels come into contact, thanks to diffusion, they can evaporate or condense. In our model, the dynamics of the air parcel is considered to be diffusive, i.e. the air parcels diffuse over the total system, experiencing the cohesive interactions. As a result of the diffusion of air parcels, the surrounding regions above the cloud may become part of the cloud by condensation, or vice versa. For this diffusion, we use a sandpile-like dynamic, in which the air parcels are transferred to the surrounding regions according to the energy content of this part and the surrounding parts. \\

We consider a two-dimensional $L\times L$ square lattice with open boundary conditions. Note that since this system is out of equilibrium, the local temperature, the specific humidity, and the local partial pressure differ from the ambient around. Therefore, we attribute to each site $i$ in our lattice a specific humidity $q_v(i)$, a local partial pressure $e_i$, and a local temperature $T_i$. Then, the cohesive energy becomes
\begin{equation}\label{eq:energy}
\text{Cohessive Energy}\equiv E_c= -J\sum_{<ij>}s_is_j-\frac{J}{2}\sum_i s_i^4, 
\end{equation} 
where $s_j=\sqrt{N_j} \in \mathbf{R}$, i.e. the square root of integer values. While this is completely different from the Ising model, due to its analogy to the Ising model, we call it the \textit{continuous Ising model (CIM)}. The sum in the first term  is only over the nearest neighbours $j$ of cite $i$.\\

We proceed further by introducing into the model the thermodynamics of mixing two different gases. Let's consider two air parcels with mass, temperature, and humidity $m_i$, $T_i$, and $q_{v}^i$, ($i=1,2$). The temperature of the mixture depends on the thermodynamics of the process, which is supposed to be adiabatic in this case. When these air parcels mix, for an isobaric process, we have
\begin{equation}
	T_{\text{mix}}=\frac{m_1T_1+m_2T_2}{m_1+m_2}.
	\label{Eq:mix}
\end{equation} 
When there are two cells, and the amount $\delta q_v$ transits from cell $i$ to cell $j$, we have 
\begin{align}
	q_v^{\text{mix}}(i) = q_v(i)-\delta q_v\\
	q_v^{\text{mix}}(j) = q_v(j)+\delta q_v.
	\label{Eq:mixingHumidity}
\end{align}
When the water vapor content of an air parcel changes by $\delta m_v$, $q_v$ in turn changes by the amount 
\begin{equation}
	\delta q_v=\frac{\delta m_v}{m_m}-\frac{m_v\delta m_m}{m_m^2}\approx \frac{\delta m_v}{m_m}, 
\end{equation}
since $\sfrac{m_v}{m_m}\ll 1$. We used the proportionality relation $\delta q_v\propto (e_i-e_j)$, justified by the fact that when water vapour is transferred from site $i$ to site $j$ their $q_v$'s change correspondingly. Using this relation, and Eq.~\ref{Eq:mix} for the isobaric mixing, it easy to find that
\begin{align}
	T_i^{\text{mix}} = &\,T_i\\
	T_j^{\text{mix}} = &\,\frac{\delta q_v}{q_v^j+\delta q_v}T_i+\frac{q_v^j}{q_v^j+\delta q_v}T_j.
	\label{Eq:mixisobar}
\end{align}
For an adiabatic process, however, one can easily show that~\cite{lohmann2016}
\begin{align}
	T_i^{\text{mix}} = &\,T_i\left(1-\gamma\frac{\delta q_v}{q_v^i} \right)\\
	T_j^{\text{mix}} = &\,T_j\left(1+\gamma\frac{\delta q_v}{q_v^j} \right),
	\label{Eq:mixadiabatic}
\end{align} 
where $\gamma=\frac{c_p}{c_v}$, and $c_p$ ($c_v$) is the isobaric (isovolume) specific heat. Furthermore, after finally reaching a mix temperature $T_{\text{mix}}$, the mixture may change its phase, e.g. from vapor to a liquid or vice versa. In this case, the temperature should be read by the coexistence saturation temperature given by Eq.~\ref{Eq:saturation}, and represented in Figure~\ref{fig:mixture}, which illustrates this situation. In this figure, the bold and dotted lines represent the temperature and pressure of the mixture and the saturation curve, respectively. A straightforward calculation shows simply that the final temperature in these situations should be
\begin{multline}
    T_{\text{new}}\equiv T'_{\text{mix}}\approx\\T_{\text{mix}}\left(1-\frac{2T_\text{mix}}{B-T_\text{mix}}-\frac{2Pc_pT_\text{mix}^3}{B\epsilon L_\nu(B-T_\text{mix})e_s(T_\text{mix})}\right),
    \label{Eq:temp}
\end{multline}
where $T_{\text{mix}}$ is the temperature of the air parcel after mixing, $T'_{\text{mix}}$ is the temperature of the mixture after changing the phase, $L_v$ is the latent heat of evaporation, and $c_p$ is the specific heat (the details of the calculation can be found in appendix \ref{app:temp}). At this point, we have the necessary prerequisites to construct our model, which is mixture of two models: the CIM, and a sandpile-like diffusion model. We employed a Monte Carlo scheme explained in the next section to handle it numerically.

\section{Our Model}

Now we describe our algorithm for simulating internal dynamics of the cumulus clouds based on a Metropolis Monte Carlo method. We consider only the dynamics of moist air in a $L\times L$ square lattice. We attribute the following quantities to each point of this lattice: the humidity $q_v^i$, the partial pressure $e_i$, the local temperature $T_i$ and the cloud field $\text{cloud($i$)}$. The latter being FALSE if the site $i$ is in the vapor phase and TRUE if it is in the condensed phase. For every site $i$ on the lattice, we have
\begin{equation}
	\text{cloud($i$)} = \begin{cases}
		\text{True}\quad \text{if}\quad q_v^i\ge q_s(T_i,P)\\
		\text{False}\quad \text{if}\quad q_v^i < q_s(T_i,P),\\
	\end{cases}
\label{Eq:cloud}
\end{equation}
 The partial pressure for the site $i$ is given the value
\begin{equation}
	e_i = \begin{cases}
		\frac{P}{\epsilon}q_s(T_i,P) &\quad\text{if}\quad \text{cloud($i$) = True}\\
		\frac{P}{\epsilon}q_v^i &\quad\text{if}\quad \text{cloud($i$) = False.}\\
	\end{cases}
\label{Eq:pressure}
\end{equation}
Also the cohesive energy of the cloud is given by the CIM: 
\begin{equation}
	E_c=-J\sum_{\langle i,j\rangle}s_is_j-\frac{J}{2} \sum_is_i^4,
	\label{Eq:Energy}
\end{equation}
where this time $s_i$ takes care of the number of water droplets in a cell, i.e. $s_i\equiv \sqrt{q_v^i}\Theta(q_v^i-q_s(T_i,P))$, where $\Theta$ is the step function. The sum over $\langle i,j\rangle$ includes only the nearest neighbors $j$ of the selected site $i$.\\

The algorithm runs as follows: We start with a prefixed pressure, temperature, and humidity of the air around the cloud, We attribute random uncorrelated humidity and temperatures to the sites of the lattice, and assume that each site contains an air parcel with uniform humidity $q_v^i$ and temperature $T_i$. The average temperature $\bar{T}$ is equal to the temperature of the surrounding air, and the specific humidity $\bar{q}_v$ is equal to $q_s=\frac{\epsilon}{P}A\exp\left[-\frac{B}{\bar{T}} \right] $ at the LCL. We randomly distribute the water particles throughout the system, such that inside each cell the specific humidity and temperature are random numbers drawn from uniform distributions as follows:
\begin{align}
	q_v^i\in &\left[\bar{q}_v-\Delta q/2,\bar{q}_v+\Delta q/2 \right]\\
	T_i\in &\left[\bar{T}-\Delta T/2,\bar{T}+\Delta T/2 \right].
\end{align}
As a result, we obtain the corresponding partial pressures ($e_i$), saturation humidity ($q_s^i$), the total energy, and the cloud field according to Eqs.~\ref{Eq:cloud},~\ref{Eq:pressure}, and~\ref{Eq:Energy}. Then we follow the following steps (schematically shown in figure \ref{fig:SchematicAlgorithm}):

\begin{enumerate}[wide, labelwidth=!, labelindent=0pt]
	\item Randomly choose a site ($i$) for the particle transfer to its nearest neighbor with minimum pressure ($j$). If $j$ is outside the lattice, then the transition takes place. Otherwise, apply the following test to measure whether the transition is permissible or not. 
	\item If it turns out that $e_i>e_j$, then there is a necessary but not sufficient condition for the number $\delta q_v=\frac{\epsilon}{2P}\left( e_i-e_j\right)$ of condensed water particles to move from site $i$ to site $j$. For this to happen, a few other conditions have to be satisfied. Determine these conditions by allowing the mixing to take place virtually and perform the following test steps:
    \begin{enumerate}
        \item After mixing (which is considered to be adiabatic), we calculate $q_v$ and $T$ for the two neighboring sites $i$ and $j$. Eq.~\ref{Eq:mixingHumidity} implies 
        \begin{align}
            q_v^{\text{test}}(x) = q_v(x)-\delta q_v^x
        \end{align}
       where $x=i,j$, $\delta q_v^i=-\delta q_v^j=\delta q_v$, and also the temperature is concluded using Eq.~\ref{Eq:mixadiabatic}. As a result, each site's constituents may change the local phase.
        
        \item Based on the result in (a), we determine the new functions Evap($i$) and Evap($j$), defined as $\text{Evap}=+1(-1)$ if the phase change is from condensed to vapor (vapor to condensed) and zero if there is no phase change. Using Eq.~\ref{Eq:cloud}, determine the virtual status of the two affected sites and also the virtual spins $s_{\text{test}}^i$ and $s_{\text{test}}^j$ so that using the Eq.~\ref{Eq:Energy} one obtains the change of the cohesive energy that showed by $\delta E_c$.
        \item Define
        \begin{equation}
        	\delta E_{x}\equiv L_v\text{Evap}\left(x\right) \left( q_v^x-\delta q_x\right)
        \end{equation}
        where $x=i,j$, $\delta q_i=-\delta q_j=\delta q$, and also 
        \begin{equation}
        	\delta E_{\text{total}}\equiv\delta E_c+\delta E_i+\delta E_j.
        \end{equation}
        Then with the probability 
        \begin{equation}
        P\equiv \text{min}\left\lbrace 1,e^{-\beta_{ij}\delta E_{\text{total}}}\right\rbrace
        \end{equation}
     the transfer takes place, where $\beta_{ij}^{-1}\equiv T_i+T_j$ is the inverse of the average local temperature.
    \end{enumerate}
    \item If the transfer of particles takes place, the test changes calculated in step (2) are accepted. In addition, the temperatures $T_{\alpha_i}$ and $T_{\alpha_j}$ of the nearest neighbours $\alpha_i$ and $\alpha_j$ of sites $i$ and $j$  change as follow:
    \begin{align}
        T_{\alpha_x}\rightarrow &T_{\alpha_x}-\frac{\delta E_x}{3c_pq_v^{\delta_x}},
    \end{align}
    and the test change of temperatures of sites $i$ and $j$ are accepted, with an additional update: if a condensation has taken place, then the temperature changes additionally according to Eq.~\ref{Eq:temp} to move the point onto the saturation humidity curve. 
	\item Go to step 1.
\end{enumerate}

In the algorithm mentioned above, the water particles are allowed to enter the system from the outside (surrounding air) into any cloud cell, just as they can leave the system. Let us call the process explained in the algorithm \textit{relaxation process}, and the entry of particles as \textit{particle injection}. After $n=10$ relaxations, one particle is injected into a random site of the lattice. \\

To clarify the algorithm, we sketch it schematically in Fig.~\ref{fig:SchematicAlgorithm}. Suppose we choose site $i$ as the point from which there is a particle transfer to the neighboring site $j$, which has the lowest pressure in the system. These two sites are colored yellow in Fig~\ref{fig:SchematicAlgorithm} and have three nearest neighbors each, colored in light blue if they belong to site $i$ and orange for site $j$. Then the amount  $\delta E_{\text{total}}$ of energy is consumed, which is why we have used the metropolis algorithm to model the change. As a secondary effect of this transition, the state of the sites $i$ and $j$ may change (condense or evaporate), and consequently, the temperatures of the nearest neighbors (light blue and orange sites in the figure) change, thus providing the required energy. In addition, the temperature of the sites $i$ and $j$ change correspondingly. The dynamics is similar to toppling sandpiles that trigger avalanches, endowed with Ising-like Monte Carlo steps for a given temperature. \\

Figure~\ref{fig:sh1} shows the evolution of the cloud in terms of Monte Carlo steps for $T=2.7$, $Q_s=3.17$, and $P=0.93 \text{atm}$, in which the yellow cells are in the condensed phase (the cloud), and blue cells are in the vapor phase. We see that the internal structure of the cloud forms as a function of Monte Carlo steps, $n$. The cloud reaches its steady state when the total energy becomes almost constant (note that the energy initially decreases with time). This steady state is shown in Fig~\ref{fig:t6}.

\section{results}
\label{result}
\begin{figure*}
	\centering
	\begin{subfigure}{0.33\textwidth}\includegraphics[width=\textwidth]{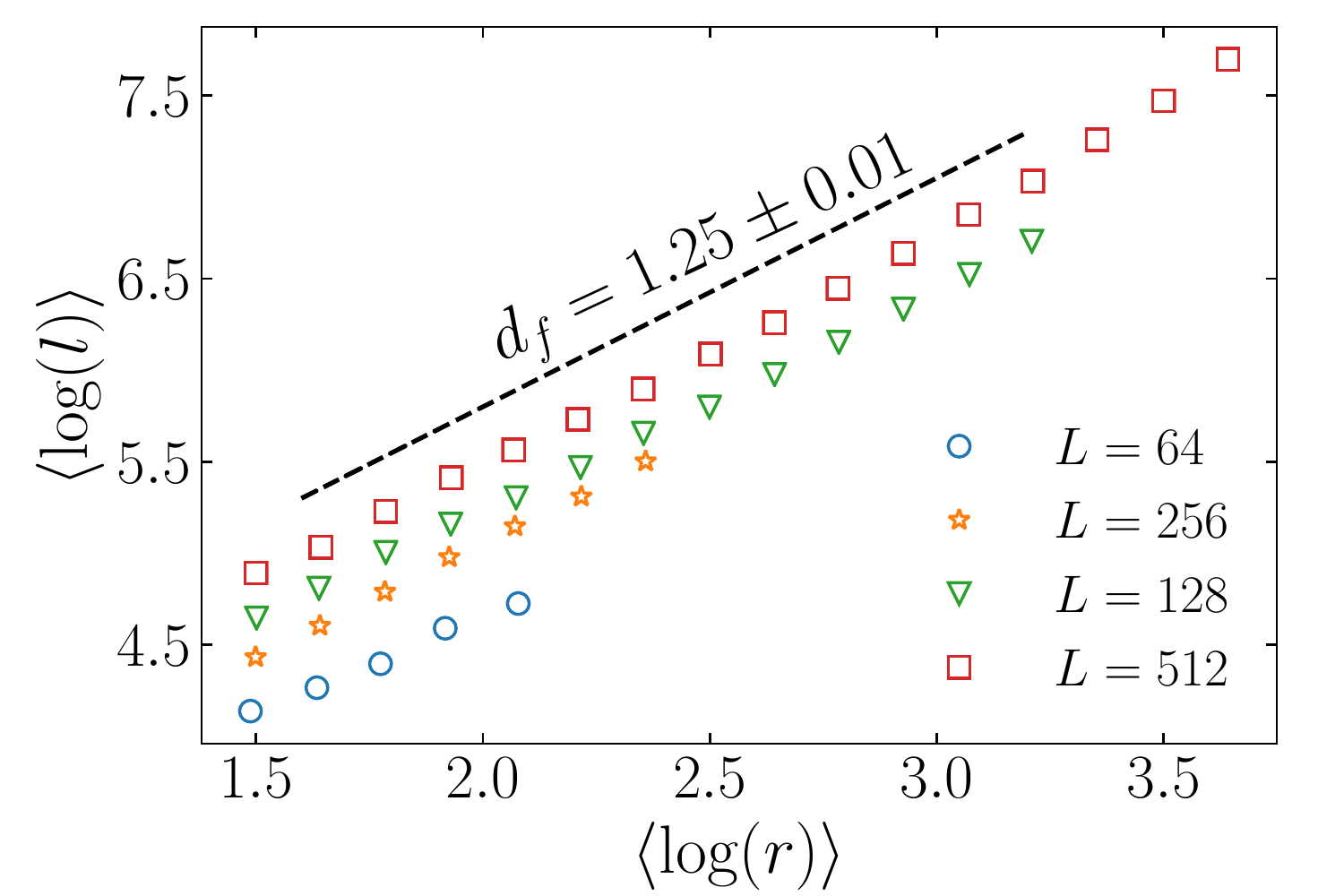}
		\caption{}
		\label{fig:simulation_FD3}
	\end{subfigure}
	\centering
	\begin{subfigure}{0.33\textwidth}\includegraphics[width=\textwidth]{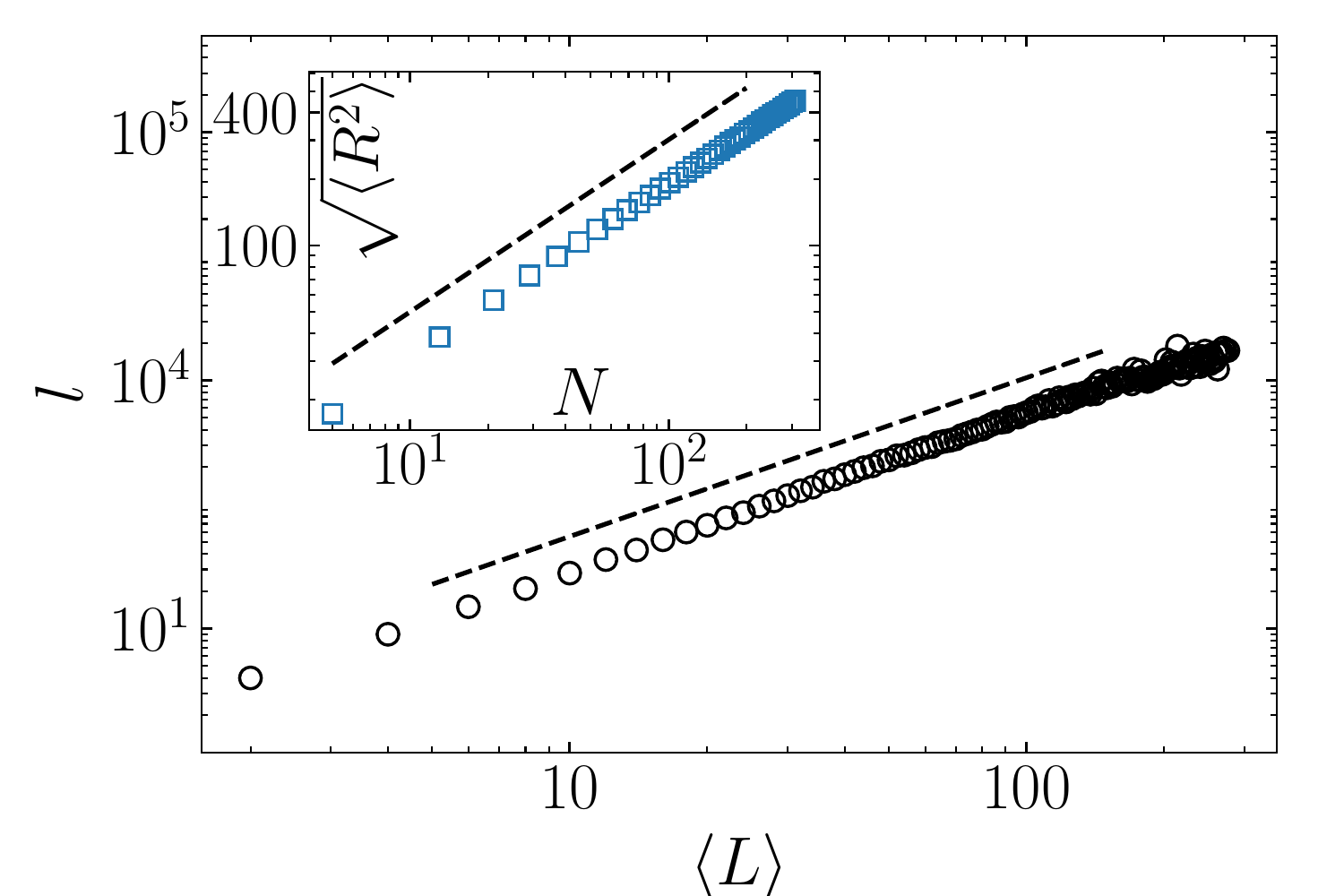}
		\caption{}
		\label{fig:simulation_FD5}
	\end{subfigure}
	\begin{subfigure}{0.32\textwidth}\includegraphics[width=\textwidth]{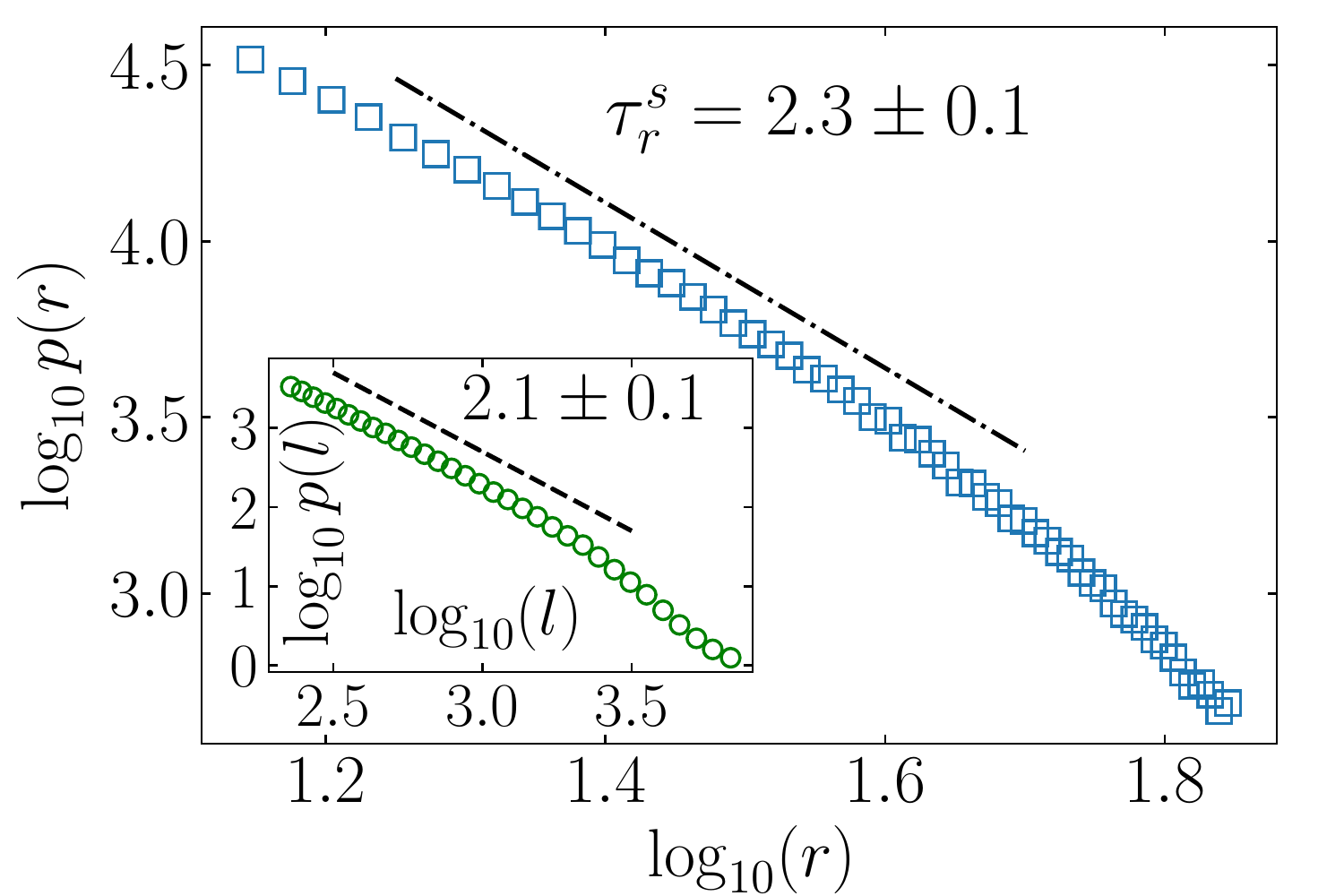}
		\caption{}
		\label{fig:p_x0}
	\end{subfigure}
	\begin{subfigure}{0.33\textwidth}\includegraphics[width=\textwidth]{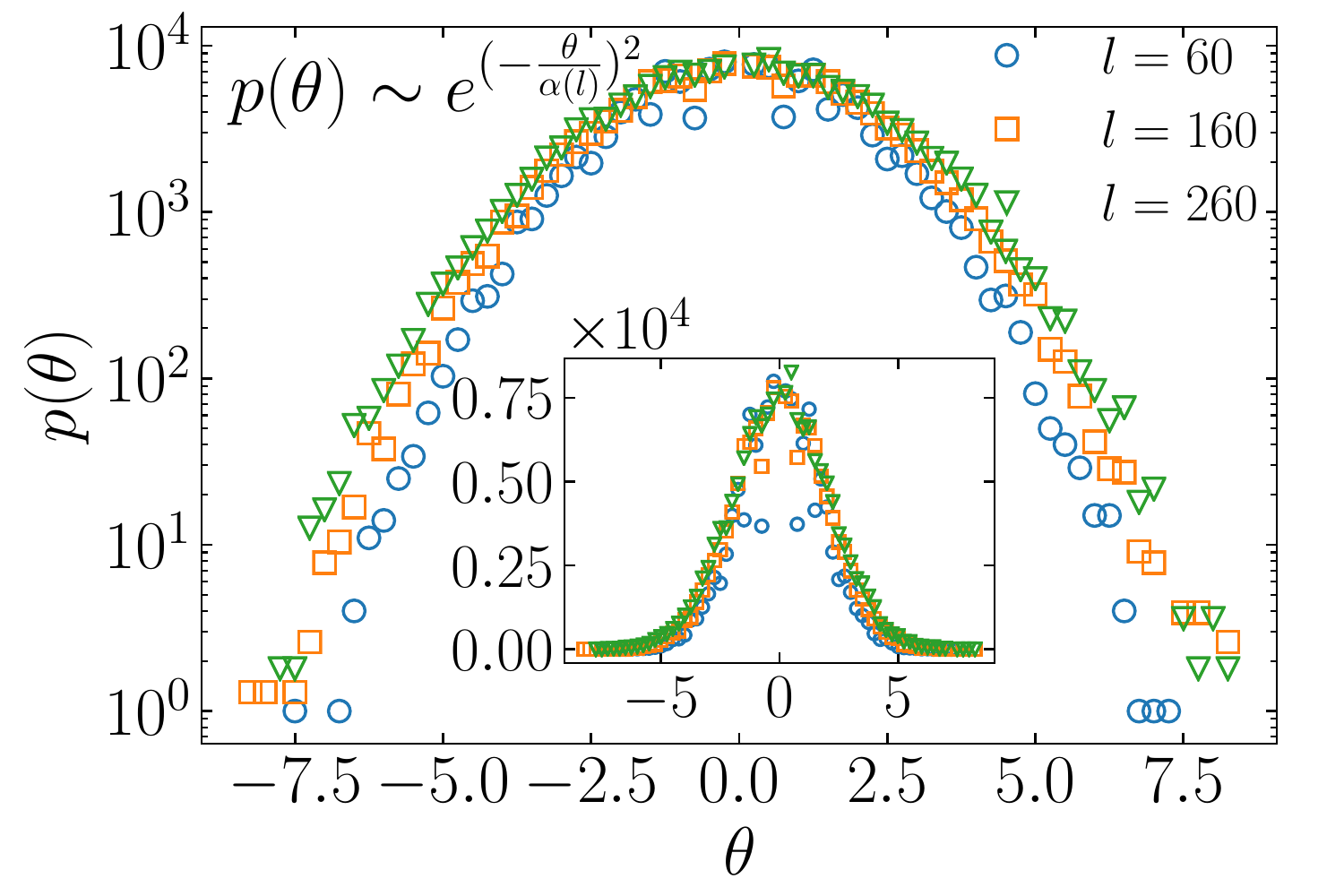}
		\caption{}
		\label{fig:sp_t_loop}
	\end{subfigure}
	\begin{subfigure}{0.33\textwidth}\includegraphics[width=\textwidth]{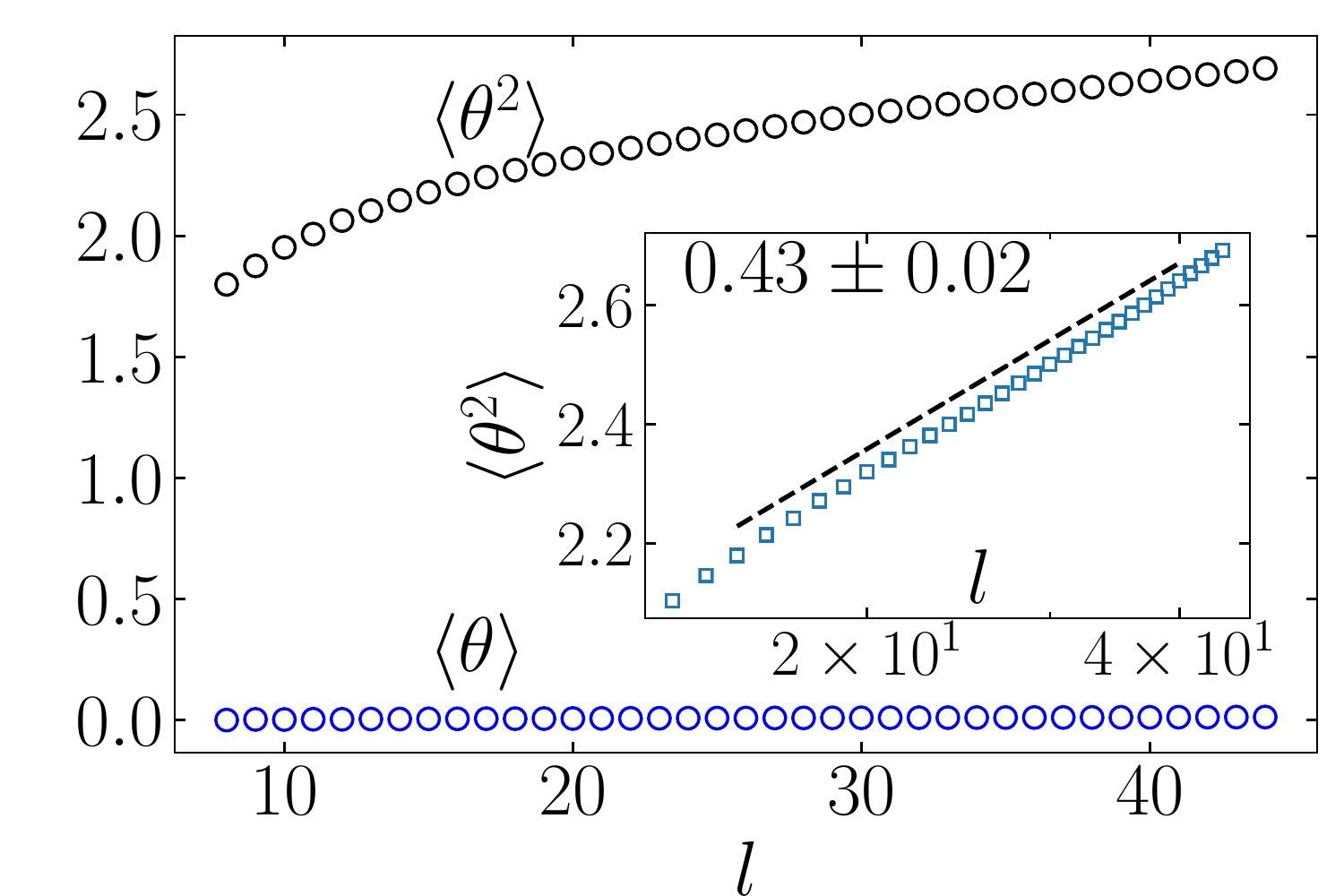}
		\caption{}
		\label{fig:sWA_loop}
	\end{subfigure}
	\begin{subfigure}{0.32\textwidth}\includegraphics[width=\textwidth]{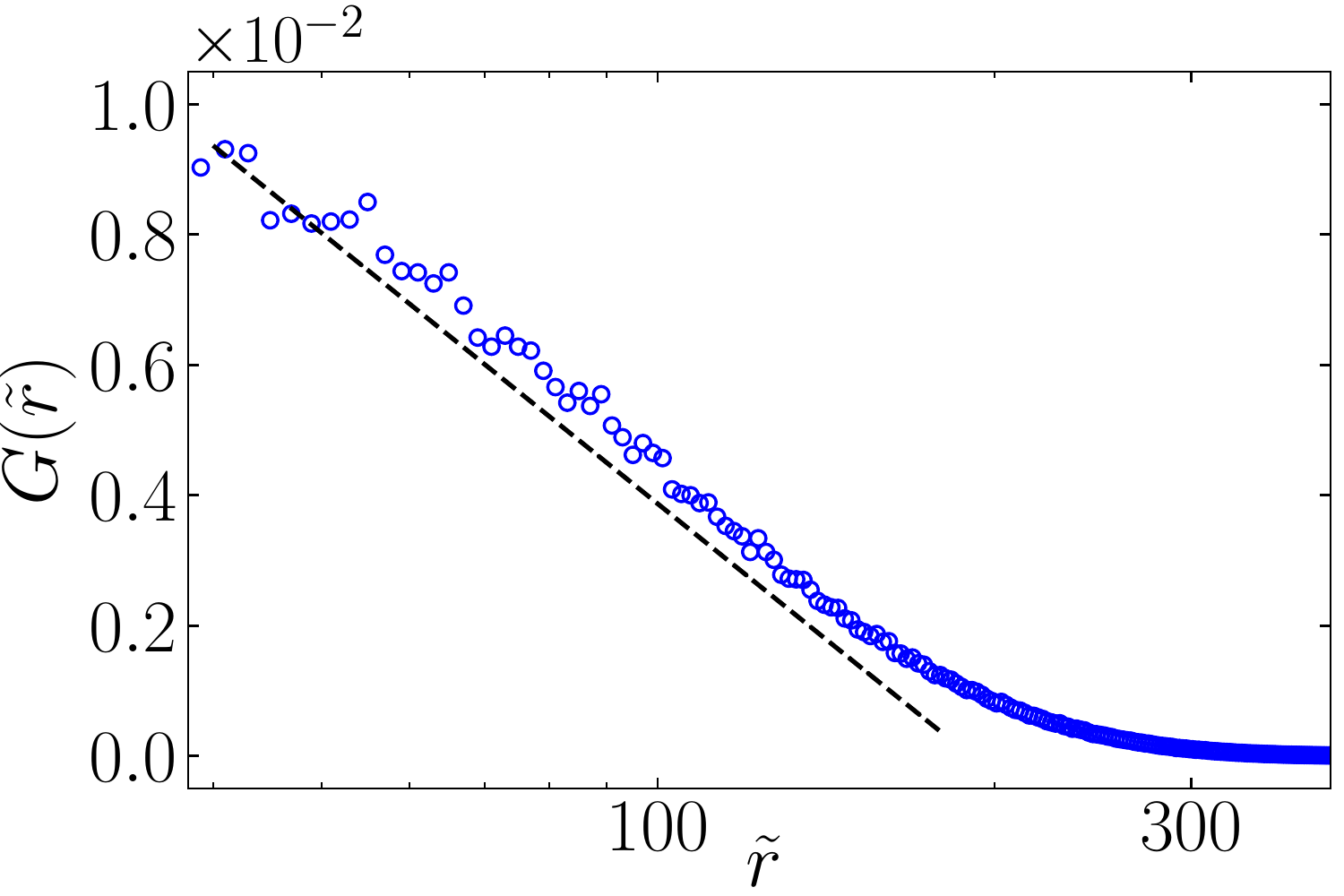}
		\caption{}
		\label{fig:G_r1}
	\end{subfigure}
	\caption{(Color online) The fractal dimension resulting from (a) $\log$-$\log$ plot of $l-r$ scaling, and (b) the sand box method from the simulations. $\nu_s$ shows the end-to-end distance exponent, which is equal to $\frac{1}{d_f}$. (c) The distribution functions of $l$ and $r$ with the exponents $\tau_r^s$, and $\tau_l^s$. (d) The distribution function of $\theta$ in the winding angle statistic which is Gaussian. (e) The variance of $\theta$ which is logarithmic in terms of $l$ with the slope $0.43\pm 0.02$. (f) The semi-logarithmic plot of the loop Green function from the simulations, that is logarithmic consistent with the 2D BTW model.}
	\label{fig:simulation_FD0}
\end{figure*}

 We report the numerical results for the model introduced in the previous section. We calculate the fractal dimension as well as the critical exponent of the distribution function of gyration radius of cloud clusters. We also determine the winding angle statistic of line levels and the loop Green's function. The fractal dimensions, $d_f$, of the interface between the vapor and condensed phase can be determined using the sand box method 
\begin{equation}
l \sim \langle L \rangle^{d_f},
\label{Eq:FD1}
\end{equation}
where $l$ is the curve length in a box of length $L$. In addition, the correlation length exponent, $\nu$, is defined by 
\begin{equation}
\langle R_N^2 \rangle  = N^{2\nu}
\label{Eq:FD2}
\end{equation}
in which $R_N$ is the end-to-end Euclidean distance between the starting point and the $N^{th}$ point of the curve. This exponent is related to the fractal dimension 
\begin{equation}
\nu = \frac{1}{d_f}.
\end{equation}
For the case we have loops, i.e. the interfaces are closed, we can use the relation
\begin{equation}
\langle \log l \rangle = d_f \langle \log r\rangle + \text{constant},
\label{Eq:FD3}
\end{equation}
where $r$ is the radius of gyration defined by
\begin{equation}
	r^2 = \frac{1}{l} \sum_{i=1}^{l} (\vec{r}_i - \vec{r}_{\text{com}}),
\end{equation}
where $\vec{r}_{\text{com}}= \frac{1}{l} \sum_{i=1}^{l} (\vec{r}_i)$ is the center of mass. Figures~\ref{fig:simulation_FD3}, and~\ref{fig:simulation_FD5} show the results of Eq.~\ref{Eq:FD3} (loop fractal dimension) and Eq.~\ref{Eq:FD1} (the results for the sand box method) and Eq.~\ref{Eq:FD2} (end-to-end distance fractal dimension), respectively. The inset demonstrates that the end-to-end distance exponent is $\nu_s=0.79\pm 0.01$. All of these results support the conclusion that $d_f=1.25\pm 0.01$ and $1.26\pm 0.01$. These results are consistent with the fractal dimension of the observational visible light in Ref~\cite{PhysRevE.103.052106}. This fractal dimension is consistent with the fractal dimension of the external frontier of the avalanches in 2D sandpile models~\cite{najafi2012avalanche,najafi2014bak,najafi2016bak,najafi2018coupling}.\\ 

The distribution function of the geometrical observables like the loop length, $l$, and the radius of gyration, $r$, was shown to follow a power-law behavior for cumulus clouds in Fig~\cite{PhysRevE.103.052106} as $P(x)\sim x^{-\tau_x}$, for $x=r$ and $l$, where $\tau_r=2.12 \pm 0.03$ and $\tau_l=2.38\pm 0.02$ for the observational data~\cite{PhysRevE.103.052106}. Figure~\ref{fig:p_x0} shows the distribution function of the $r$ and $l$. The corresponding exponents obtained by simulating our model are $\tau_r=2.3\pm 0.1$ and $\tau_l=2.1\pm 0.1$. These results are again consistent with the observational data.\\

The winding angle statistic is based on the variance of the angle between the straight line joining two selected points (origin point and end point) on the curve with distance $l$ and the tangent line to the curve at the origin point, which is defined as follows:
\begin{equation}
	Var[\theta] = C+2\frac{d_f-1}{d_f} \ln l,
	\label{var}
\end{equation}
where $Var[\theta]$ is the variance of angle ($\theta$) and $C$ is an arbitrary constant. Fig.~\ref{fig:sp_t_loop} shows the distribution function of $\theta$, which indicates the fact that the distribution function is Gaussian, and Fig.~\ref{fig:sWA_loop} reveals that the variance of $\theta$ is logarithmic with $l$, with a slope $0.43\pm 0.02$. This slope corresponds to the fractal dimension of $1.27\pm 0.02$. Therefore, we confirm that the statistical observables considered are consistent with each other and also with the observational data in Ref.~\cite{PhysRevE.103.052106}. One can compare the results of our method and those of Ref.\cite{PhysRevE.103.052106} reported in table \ref{tab:conditions}, which includes simulation and observational data.

\begin{table*}
	\begin{tabular}{|c | c | c | c|c|c|c| }
		\hline method & $\kappa$ & $d_f$ & $2(d_f-1)/d_f$ & $\nu$&$\tau_r$ & $\tau_l$\\
		\hline Observational data & $2.1\pm 0.2$ & $1.248\pm 0.006$ & $0.42 \pm 0.01$ & $0.81 \pm 0.01$& $2.38 \pm 0.02$&$2.12 \pm 0.03$ \\
		\hline CML simulation & $2.024\pm 0.06$ & $1.247 \pm 0.016$ & $0.40 \pm 0.05$&  -&$2.35 \pm 0.06$&$2.14 \pm 0.05$\\
		\hline ALG1 simulation  & $2.0 \pm 0.1$ & $1.25 \pm 0.01$ & $0.43 \pm 0.04$& $0.79 \pm 0.01$&$2.3 \pm 0.1$&$2.1 \pm 0.1$ \\
		\hline 2D BTW & $2$ & $1.25$ & $0.4$&-&$\frac{5}{3}$&$1.28$  \\
		\hline
	\end{tabular}
	\caption{\label{tab:conditions}The numerical values of the diffusivity parameter $\kappa$ and the fractal dimension $D_f$ obtained by various methods. The data for the ordinary 2D BTW model is also shown for comparison (although $\tau_r$ and $\tau_l$ are different).}
\end{table*}

Schramm-Loewner evolution (SLE) is a powerful tool for analyzing the critical behavior of two-dimensional systems through loop-less curves (in the cloud lattice system, these would be the boundaries separating the condensed phase from the gas phase). SLE can classify these curves into single parameter classes via a diffusive parameter $\kappa$. Furthermore, this parameter can be related to the system's fractal dimension through the relation $d_f = 1+ \frac{\kappa}{8}$ \cite{cardy2005sle}.  Given the profound connection between SLE and conformal field theory (CFT), one can classify conformally invariant models into CFT universality classes. In that case, the parameter $\kappa$ is related to the central charge $c$ of the conformal field theory by $c = \frac{(8-3\kappa)(\kappa-6)}{2\kappa}$ \cite{rohde2011basic,cardy2005sle}. According to these facts, our model could be considered conformal symmetry compatible with $\kappa=  2 $, which leads to the $c = -2$ CFT universality class. Therefore, the model falls into the same universality class as the Loop Erased Random Walk (LERW) model and the cloud-air boundary of real cumulus clouds whose $\kappa$ was determined to be equal to 2 \cite{najafi2012avalanche,PhysRevE.103.052106}.\\

The other important quantity is the loop Green's function, $G(\tilde{r})$, defined as the probability that two points at a distance $\tilde{r}$ belong to the same closed interface. It is logarithmic for visible light arriving from clouds~\cite{PhysRevE.103.052106}, and the results related to our model are shown in Fig.~\ref{fig:G_r1}, where It reveals that the loop Green's function is logarithmic.

\section{conclusion}
In this study, using the fact that specific humidity, temperature, pressure, and the cohesive energy of water droplets play a substantial role in cloud formation, we introduced a new model for the formation of cumulus clouds in LCL. In this algorithm, based on Monte Carlo, we included many physical processes like the evaporation and condensation, the cohesive energy between water droplets, the diffusion, the heat releasing due to the phase transitions, and the thermodynamics of mixing of air parcels. The \textit{separation of time scales} and the \textit{energy threshold for toppling} as two building blocks of self-organized critical (SOC) systems is naturally present in our model, consistent with the previous observations. Our model consists of two ingredients: sandpile-like topplings that are responsible for the diffusion process of air parcels, and a continuous Ising-like Hamiltonian responsible for cohesive energy between water droplets. The simulation result reveals that the algorithm works well and that its exponents show the same result as observational visible light of cumulus clouds results and CML method simulation. Particularly, we show that the fractal dimension of the cloud clusters is $d_f=1.25\pm 0.01$, consistent with 2D sandpiles, and also the Green function turns out to be logarithmic with distance. The exponents of the distribution function of the loop length and the radius of gyration, although different from the sandpiles, are consistent with the observational data. 

\bibliography{refs}

\appendix

\section{Temperature at saturation}\label{app:temp}
The slope of the \textbf{isobaric} condensation line connecting the points $(T_{mix},e_{mix})$ and $(T'_{mix}, e'_{mix})$ in figure \ref{fig:mixture} is given by:
\begin{equation}\label{eq:vapor_pressure_derivative}
    \frac{de}{dT}=\frac{p}{\epsilon}\frac{dq_\nu}{dT}=-\frac{pc_p}{\epsilon L_\nu},
\end{equation}
where $e$ is the partial pressure of water vapor in moist air, which is often used to characterize the water vapour content in the atmosphere. The intersection of this line with the saturation vapor pressure curve (dashed curve in figure \ref{fig:mixture}),
\begin{equation}\label{eq:sat_vapor_pressure}
    e_s(T') =A\exp(-B/T'),
\end{equation}
i.e. $e(T')=e_s(T')$ determines the resulting temperature $T'_{mix}$ and vapor pressure $e'_{mix}$ after the condensation process has terminated \cite{lohmann2016}. We integrate equation \ref{eq:vapor_pressure_derivative} between the starting temperature $T$ and the final temperature $T'$ that we want to determine. The integral is given by
\begin{equation}\label{eq:vapor_pressure}
    e(T') = e(T) + (T-T')\frac{c_pP}{\epsilon L_\nu}.
\end{equation}
Equating  \ref{eq:vapor_pressure} and \ref{eq:sat_vapor_pressure} we obtain the following equation for $T'$.
\begin{equation}\label{eq:intersection}
    e(T)+(T-T')\frac{c_pP}{\epsilon L_\nu}=A\exp(-B/T').
\end{equation}
By expanding $e_s$ with $T'=T-\delta T$ we have
\begin{equation}
    e_s(T-\delta T)\approx e_s(T)-\delta Te_s'(T)+\frac{1}{2}\delta T^2e_s''(T).
\end{equation}
where,
\begin{align}
    e_s'(T) =& \frac{B}{T^2}e_s(T), \label{eq:e_s'}\\
    e_s''=&\frac{B}{T^3}\left(\frac{B}{T}-1\right)e_s(T).\label{eq:e_s''}
\end{align}
By inserting the expansion into \ref{eq:intersection} and using the approximation $e(T)\approx e_s(T)$, we obtain
\begin{equation}
    \frac{c_pP}{\epsilon L_\nu}\equiv m \approx -e_s'(T)+\frac{1}{2}\delta Te_s''(T),
\end{equation}
so that 
\begin{equation}\label{eq:deltaT}
    \delta T = T-T'=2\frac{m+e_s'(T)}{e_s''(T)},
\end{equation}
where $e_s''(T)\neq 0$. Finally, after substituting \ref{eq:e_s'} and \ref{eq:e_s''} into \ref{eq:deltaT} we obtain
\begin{equation}\label{eq:T'}
    T'=T\left(1-\frac{2T}{B-T}-\frac{2Pc_p}{B\epsilon L_\nu(B-T)e_s(T)}T^3\right).
\end{equation}
\end{document}